\documentclass[journal=jacsat,manuscript=article]{achemso}

\usepackage[version=3]{mhchem} 
\usepackage{natbib}

\usepackage[dvipsnames, table]{xcolor} 
\usepackage{multirow} 
\usepackage{graphicx} 
\usepackage{adjustbox} 
\usepackage{makecell} 
\usepackage{amssymb} 
\usepackage{array} 
\usepackage[utf8]{inputenc} 
\usepackage{boldline}
\usepackage{xr}
\usepackage[hidelinks]{hyperref}

\makeatletter
\newcommand*{\addFileDependency}[1]{
\typeout{(#1)}
\@addtofilelist{#1}
\IfFileExists{#1}{}{\typeout{No file #1.}}
}\makeatother

\newcolumntype{P}[1]{>{\centering\arraybackslash}p{#1}}

\DeclareUnicodeCharacter{2009}{\,}

\usepackage{placeins} 

\usepackage[bottom]{footmisc}

\author{Wai H. Oo}
\affiliation{Department of Mechanical Engineering, University of California Merced, Merced, California 95343, USA}
\altaffiliation{These authors contributed equally to this work}

\author{Hongyu Gao}
\affiliation{%
Department of Materials Science and Engineering, Saarland University, Saarbrücken, Saarland 66123, Germany
}%
\altaffiliation{These authors contributed equally to this work}
\email{hongyu.gao@uni-saarland.de}

\author{Martin H. M\"user}
\affiliation{%
Department of Materials Science and Engineering, Saarland University, Saarbrücken, Saarland 66123, Germany
}%

\author{Mehmet Z. Baykara}
\affiliation{Department of Mechanical Engineering, University of California Merced, Merced, California 95343, USA}
\email{mehmet.baykara@ucmerced.edu}

\title[Structural Lubricity and Molecular Contamination: Rejuvenation, Aging, and Friction Switches]
  {Structural Lubricity and Molecular Contamination: Rejuvenation, Aging, and Friction Switches}

\keywords{atomic force microscopy, friction, molecular contaminants, molecular dynamics, nanotribology, superlubricity}

\begin{document}







\begin{abstract}
Using atomic force microscopy experiments and molecular dynamics simulations of gold nanoislands on graphite, we investigate why ultra-small friction commonly associated with structural lubricity can be observed even under ambient conditions. 
%
Measurements conducted within a few days after sample synthesis reveal previously undiscovered phenomena in structurally lubric systems:
\textit{rejuvenation}, a drop in kinetic friction of 
an order of magnitude shortly after the onset of sliding;
\textit{aging}, a significant increase in kinetic friction forces after a rest period of 30 minutes or more;
\textit{switches}, spontaneous jumps between distinct friction branches.
These three effects are drastically suppressed a few weeks later. 
%
Imaging of a contamination layer and simulations provide a consistent picture of how single- and double-layer contamination underneath the gold nanoislands as well as contamination surrounding the nanoislands affect structural lubricity but not lead to its breakdown.
\end{abstract}


Research on structural lubricity, an ultra-low friction state arising due to the systematic annihilation of lateral forces in atomically flat interfaces formed by two incommensurate surfaces~\cite{hirano_atomistic_1990, shinjo_superlubric_1993, muser_structural_2004}, is accelerating in recent years \cite{baykara_emerging_2018,Berman2018ACSNano}.
Although the concept came about as a theoretical exercise~\cite{hirano_atomistic_1990, shinjo_superlubric_1993}, 
atomic-force-microscopy (AFM) experiments on the friction between a graphene flake and graphite under dry nitrogen atmosphere confirmed the geometrical explanation of miniscule friction due to rotation-induced lattice mismatch~\cite{dienwiebel_superlubricity_2004}. 
%
%
More recently, other observations of ultra-low friction attributed to structural lubricity between carbon-based materials have been reported \cite{liu_observation_2012,zhang_superlubricity_2013}, complemented by reports of structural lubricity at hetero-interfaces formed by two-dimensional materials \cite{song_robust_2018, LongNL}. 
Likewise, theoretical predictions of surface contamination leading to a breakdown of structural lubricity~\cite{he_adsorbed_1999} were supported through AFM experiments performed on antimony nanoislands\footnote{For simplicity, `island' will be used in place of `nanoisland' hereafter.

.} on gold, contrasting UHV with ambient conditions~\cite{dietzel_frictional_2008}.
Remarkably, similar experiments performed on noble metal islands on graphite demonstrated that structural lubricity is not restricted to the pristine UHV environment.
Comparable values of friction force and the sub-linear scaling of friction force as a function of contact area were also observed under uncontrolled ambient conditions \cite{cihan_structural_2016, ozogul_structural_2017}. 
This leaves open the questions of when contaminants destroy structural lubricity and how they affect friction otherwise, which is relevant for the potential exploitation of superlubricity outside of UHV chambers. 

Contaminating interfacial particles were expected to destroy superlubricity because their mobility allows them to adopt positions, which are  energy minima of both surfaces simultaneously, whereby the surfaces interlock~\cite{he_adsorbed_1999}.
One reason why this argument may not always hold is that extremely smooth surfaces like that of graphite do not provide energy barriers large enough to substantially counteract sliding of an adsorbed or rather ``between-sorbed'' layer.
Another one is that organic molecules adsorbed on graphite form highly ordered domains.~\cite{rabe1991commensurability} 
They might act similarly to two-dimensional solids, whereby individual molecules have an additional constraint in the presence of another surface, which reestablishes, to a significant degree, the systematic annihilation of lateral forces at pristine interfaces formed by smooth, incommensurate surfaces.
Which effect dominates might depend on the structure and chemical nature of the counterface to graphite. 

To illuminate the role of contamination in superlubricity, we present results of AFM-based sliding (i.e., nanomanipulation) experiments on gold islands.
In contrast to other experiments on this system, we chose the tip-on-top\cite{dietzel_transition_2009} rather than the previously used and easier-to-implement push-from-the-side approach,~\cite{cihan_structural_2016, ozogul_structural_2017} because it allows islands to be dragged back and forth under a controlled load in a controlled direction, with controlled speeds.~
We also report results of molecular dynamics (MD) simulations mimicking the experiments, albeit using smaller islands and nine-order-of-magnitude larger  velocities.
Thus, direct quantitative comparisons are not possible and would remain questionable even when using scaling arguments. 
Nonetheless, relative trends of how contamination affects friction can certainly be contrasted, whereby MD can offer possible explanations for experimental observations.

\begin{figure}
\centering
\includegraphics[width=1\textwidth]{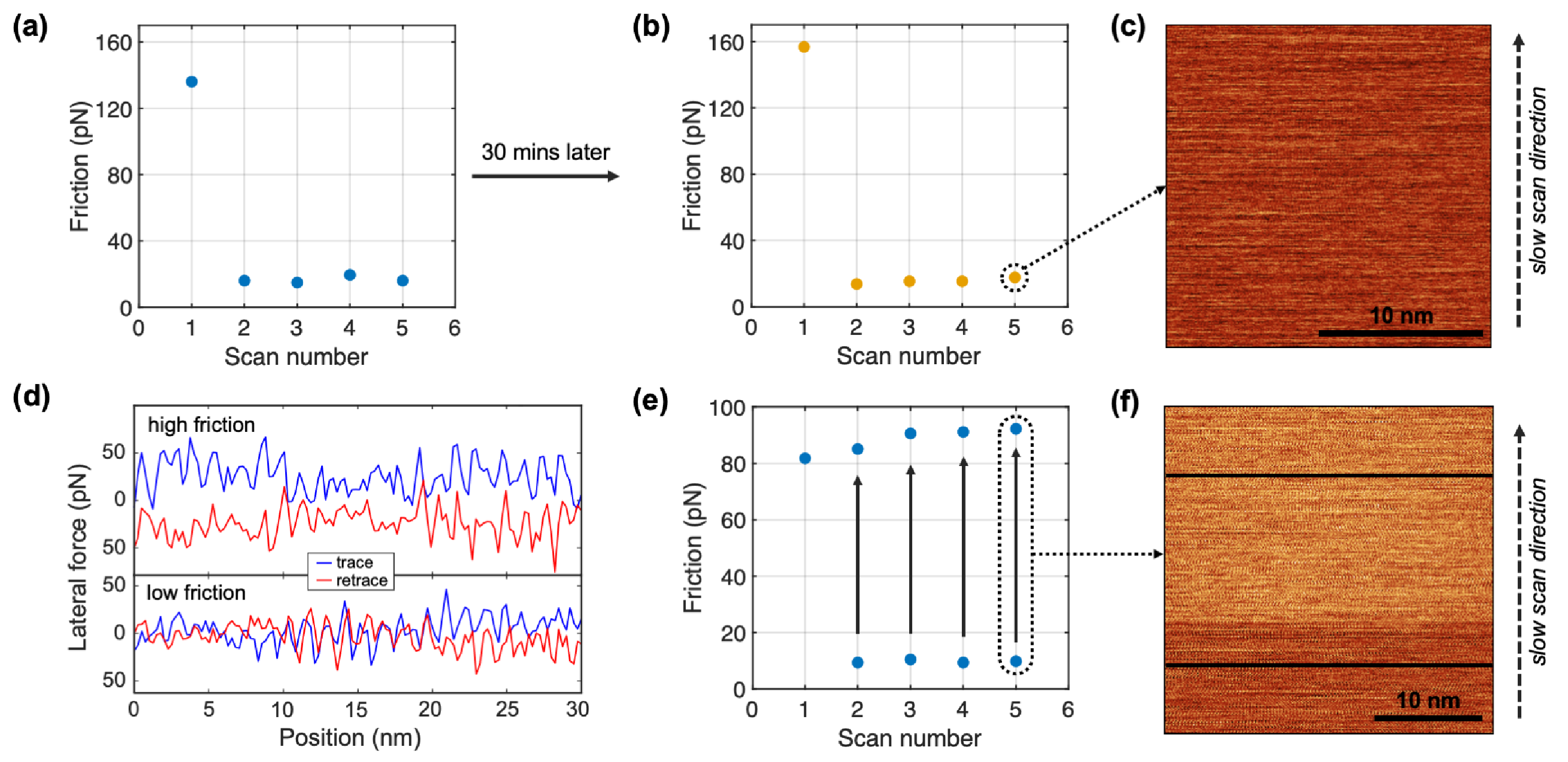}
\caption{(a) Friction force as a function of scan number extracted from the nanomanipulation of a freshly deposited island. (b) Same as (a) but repeated on the same island and area after a 30~minute waiting time.
(c) Friction-force map associated with scan~5 in (b). 
(d) Lateral forces extracted along single scan lines in the forwards and backwards directions, along the two traces shown with black lines in (f), located in high and low friction domains, respectively.
(e) Friction force as a function of scan number extracted from the nanomanipulation of another gold island. 
During the scan, friction jumped from low to high as shown in the friction-force map of panel (f). 
}
\label{fig_alleffects}
\end{figure}

Before presenting the results of the nanomanipulation experiments regarding friction forces, we first briefly describe the associated data acquisition procedure.
First, individual line scans of lateral force, such as those shown in Figure~\ref{fig_alleffects}d are recorded.
In accordance with established procedures,~\cite{Schwarz1996RSI} half of the difference between trace (i.e., forward) and retrace (i.e., backward) scans are plotted for all lines (\textit{n} = 1, 2, ..., 256), yielding maps of friction like those shown in Figure~\ref{fig_alleffects}c,f.
While some friction maps are homogeneous (Figure~\ref{fig_alleffects}c), some reveal two clearly distinct domains of high and low friction, as indicated by bright and light colors, respectively (Figure~\ref{fig_alleffects}f). 
Data points in the remaining panels of Figure~\ref{fig_alleffects} represent the average over such friction domains.
Since no scan produced more than two domains, one scan can result in one (Figure~\ref{fig_alleffects}a,b) or two (Figure~\ref{fig_alleffects}e) friction values. 
Arrows next to friction maps (Figure~\ref{fig_alleffects}c,f) indicate the slow scan direction, while arrows in the plot shown in Figure~\ref{fig_alleffects}e indicate the jumps from low to high friction domains.

Gold islands,
when measured within a few days after synthesis, always  produce relatively high friction during the first (areal) scan. 
However, friction drops substantially during the second scan, i.e., by one order of magnitude and can {remain} low in all subsequent scans, as is depicted in Figure~\ref{fig_alleffects}a.
When repeating the experiments on the same island
and same scan area,
after waiting times ranging from 30 minutes to 18 hours, similar results were obtained (Figure~\ref{fig_alleffects}b).
Differences between the average friction of the very initial scan and the first scan after $\ge 30$~minute waiting times were on the order of 15\%.
The average friction of subsequent scans were again reduced by about an order of magnitude with respect to the first one as in the initial set of experiments. 
Interestingly, stiction peaks were observed 
neither at the onset of a first scan (initial and after waiting) nor after the periodic pauses (on the order of 10 ms) at the end of scan lines, where the
sliding direction changes.
%

An increase of friction force with waiting time has been known since Coulomb's classical friction experiments~\cite{Coulomb1785MMPAR} and is commonly referred to as \textit{aging}~\cite{Li2011N}, while the reduction of friction during sliding is called \textit{rejuvenation}~\cite{Bureau2002EPJE} and captured in rate-and-state models of friction.~\cite{Dieterich1994PAGP,Chen2016JGR}
However, typical results reveal changes on the order of 10\% rather than by a factor of ten. 
Prominent examples are rough glassy polymer sliding past smooth silanized glass with nominal contact areas of order millimeter~\cite{Bureau2002EPJE} as well as material systems meant to mimic the friction of tectonic plates~\cite{Dieterich1994PAGP,Chen2016JGR}.
Given the dramatic difference in the observed effect, common explanations of aging and rejuvenation~\cite{Mueser2003ACP} do not seem plausible for our experiments. 
This suggests a rather large gap of understanding of the physical mechanisms underlying our observations at the small nanoscale.

In addition to rejuvenation and aging, we encountered “friction switches", i.e., 
spontaneous \textit{jumps} between low and high friction forces within a single scan, (see Figure~\ref{fig_alleffects}e,f).
A combination of all three effects is frequently observed within the same experiment, even with the same island on the same area (see Figure~S5).
Since graphite has a homogeneous surface away from steps and operating conditions are mild, surface contamination appears to be the most likely candidate to cause aging, rejuvenation, and friction switches in our system, which is why it is investigated next.

\begin{figure} [ht]
\centering
\includegraphics[width=0.45\textwidth]{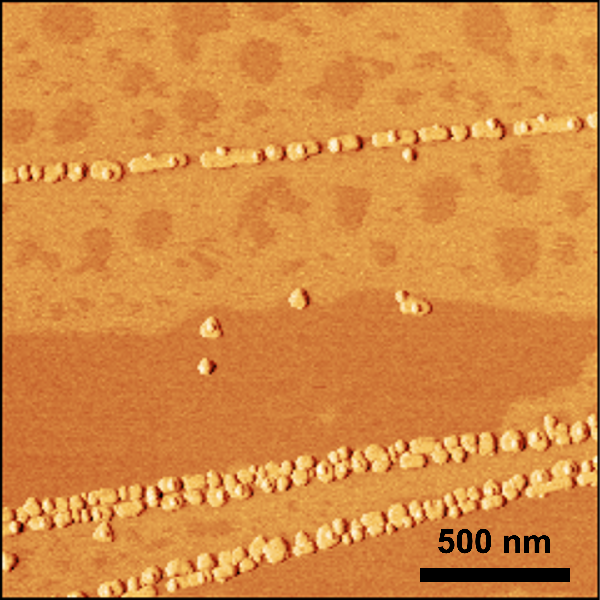}~~~~~~
\caption{AFM phase image of the sample system, taken 4 months after synthesis, showing the accumulation of contaminant layers.}
\label{fig_dirt}
\end{figure}

To gain information on contamination of our samples, we used AFM-based topographical and phase imaging.
As opposed to topographical imaging which solely provides height information, phase imaging allows materials with different mechanical stiffness on the sample surface to be distinguished with high spatial resolution.~\cite{garcia_amplitude_2010}. 
Figure~\ref{fig_dirt} shows a large-scale phase image taken via tapping-mode AFM several months after synthesis.
It reveals dark regions indicating soft contaminants and light regions for graphite and gold islands. 
Surface contamination becomes apparent as early as one week post-synthesis, despite the samples having been stored in a vacuum desiccator with the intention to deter environmental adsorbates from accumulating on the surface. 
The measured height associated with contamination, as inferred from topographical AFM imaging (see Figure~S6) ranges from a few Å up to 10~Å, which corresponds to 1--2 adsorption layers.
%


%
Manipulation of an island on a heavily contaminated sample causes a localized change in the substance's coverage (Figure~S6a,b).
However, the contamination layer can also change over time in both coverage and morphology without nanomanipulation (Figure~S6d).
Upon heating the sample to 100°C, 
phase-images have no more contrast (see Figure~S7) implying that the contaminant is desorbed or has become very homogeneous, e.g., in the form of an alkane layer with alkanes having more than $\approx 20$ carbon atoms~\cite{Paserba2001PRL,Gellman2002JPCB}.
%
According to these characteristics, the initial contaminant layer is most plausibly a mixture of water and hydrocarbons.  
As revealed by a recent study, this combination is frequently found under ambient conditions on van der Waals materials, such as graphite and hBN~\cite{palinkas_composition_2022}. 
More specifically, it is proposed that the contaminant layer comprises predominantly mid-length linear alkanes containing 20-26 carbon atoms.

Although thicker lubrication layers imply a reduced resistance to sliding in continuum mechanics and despite much knowledge about the slip of alkanes past graphitic~\cite{Falk2012L} and gold surfaces alike~\cite{Jabbarzadeh2011TI} even under high-confinement conditions~\cite{Codrignani2023SA}, friction in our system appears impossible to predict from the existing literature.
This is because friction at the nanoscale often is a true system property~\cite{Mueser2003ACP}, which cannot be deduced from those of the lubricant and its two boundaries with the confining walls. 
Details like the relative orientation of the confining walls can matter~\cite{he_adsorbed_1999} or confinement-induced structural transformation of the lubricant can occur, as might be the case for alkanes between gold~\cite{Jabbarzadeh2011TI}.
This motivated us to elucidate the dependence of shear stress on coverage and other parameters for this particular system using molecular dynamics.

\begin{figure}[ht]
\centering
\includegraphics[width=0.45\linewidth]{stress_npoly.eps}
\includegraphics[width=0.5\linewidth]{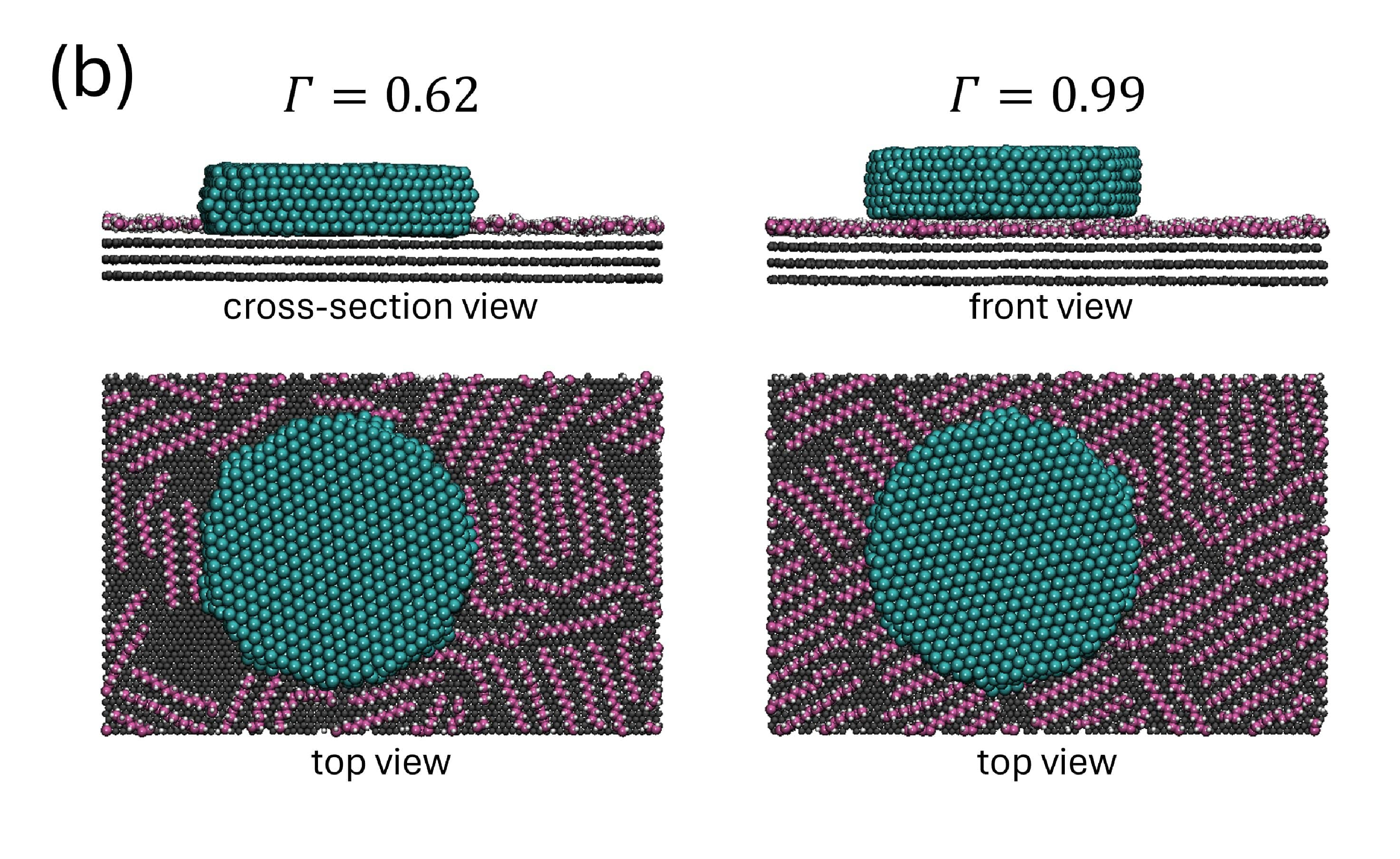}
\caption{(a) Effective shear stress ($\tau_\textrm{eff}$) as a function of the nominal coverage ($\Gamma$). Black, dashed lines separate the cases where gold slides directly on graphite at small $\Gamma$, on top of a monolayer, and on top of a double-layer of $n$-hexadecane (HEX) molecules. The blue, dashed line represents the mean shear stress for clean gold-graphite contacts. Solid and hollow symbols represent results from $v = 20$ and 2 m/s sliding velocity, respectively. (b) Snapshots at different ${\Gamma}$ at $v = 2$~m/s.}
\label{fig:stress_npoly}
\end{figure}

~
The simulations reveal a highly non-monotonic dependence of friction on coverage between the simulated island and the graphite substrate, see Figure~\ref{fig:stress_npoly}.
It is almost immeasurably small at zero coverage due to the absence of any significant instabilities~\cite{Gao2022FC}, as indicated by the blue, dashed line.
The effective shear stress, i.e., the ratio of friction force and island area, is largest at sub-mono-layer coverages ($\Gamma$) when the gold island rests directly on graphite but has to  displace polymers in direct contact with graphite out of the way while advancing.
Thus, in this case, friction stems predominantly from the out-of-contact rather than the contact region itself.
Friction drops discontinuously upon increasing $\Gamma$, once the island no longer touches graphite directly, see Figure~\ref{fig:stress_npoly}b.
It also reveals the alkanes to orient along graphite's symmetry axis, thereby forming locally commensurate domains, which explains why slip at sub-coverage occurs between gold and adsorbate.
As long as the island keeps sliding on a monolayer, friction increases with $\Gamma$ but decreases again discontinuously as soon as a second layer appears between gold and graphite.

At $\Gamma \approx 0.62$ and $\Gamma \approx 1$, simulations were run with the velocity reduced by a factor of ten.
The resulting friction drop was by a factor of {five and two}, respectively.
These changes are too significant for the friction to be labeled as Coulombic, i.e., barely dependent on velocity, although the velocities exceed the experimental ones by {nine} orders of magnitude.
It is worth noting that the trailing edge of a $\Gamma = 0.62$ contact drags along some polymers at $v = 2$~m/s but not at $v = 20$~m/s. 
Without this effect, the friction-velocity relation would have moved closer to a linear, i.e., Stokesian dependence. 
%

\begin{figure}[ht]
\centering
\includegraphics[width=0.5\linewidth]{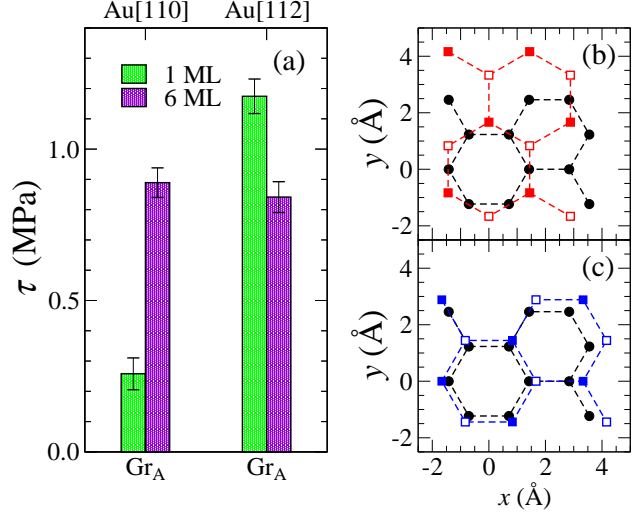}
\caption{{(a) Shear stress ($\tau$) between two area-filling graphite and gold solids separated by either a single- (1~ML) or a six-monolayer (6-ML) HEX film. 
The top and bottom ticks indicate the slab material and lattice orientation with the armchair direction of graphite (Gr$_\textrm{A}$) being aligned with the sliding velocity $\mathbf{v} = 10~\mathbf{e}_x$~m/s. 
In-plane projections of (b) Au$[1\overline{1}0]$ (red squares) and (c) Au$[11\overline{2}]$ (blue squares) on Gr$_{\rm A}$ (black circles). Solid and hollow squares indicate Au atom positions in two adjacent (111) layers, {respectively}.}}
\label{fig:stress_lattice}
\end{figure}

Since the importance of the area below the island relative to that of the contact line or circumference is larger in the experiments than in the simulations (roughly by a factor of ten, because the experimental radii exceed the simulated ones by that factor), we also simulated selected area-filling 
contacts between gold and graphite (Figure~\ref{fig:stress_lattice}).
For this purpose, we explored two coverages, $\Gamma = 1$ and $\Gamma = 6$, as well as two relative orientations between the graphite and the Au $(111)$ surfaces so that the Au surface is ideally aligned with graphite, i.e., $[1\bar{1}0]$ is aligned with the graphite armchair direction, Gr$_\textrm{A}$, albeit with a lattice mismatch of roughly {17\%} and one time misoriented by 90$^\circ$ so that $[11\bar{2}]$ is parallel to Gr$_\textrm{A}$.
For both orientations, the true areal shear stress at $\Gamma = 1$ is rather small compared to the effective shear stress of $\lesssim 6$~MPa for the island, i.e., 0.26~MPa for the $[1\bar{1}0]$ orientation and 1.2~MPa for $[11\bar{2}]$.
At $\Gamma = 6$, shear stress changes to $\tau \approx 0.87$~MPa.
Results are shown in Figure~\ref{fig:stress_lattice}a. 

These results are interesting for a variety of reasons.
First, they show that relative orientation can matter even between (quasi-) incommensurate solids as $\tau$ differs by a factor of almost five between the two configurations at $\Gamma = 1$.
Thus, for any system, in which $\Gamma$ is small and areal friction dominates, large scatter in the data is unavoidable.
%
%
Second, the concept of continuum mechanics is far from being applicable in our system, as a six-fold thicker lubricant can yield a three-fold increase in friction, at least for the misaligned surfaces, while continuum-based concepts, in particular those based on Reynolds' thin-film equation, would expect it to drop with increasing $\Gamma$.
Third and perhaps most importantly for our present purposes, friction for the simulated islands is indeed dominated by processes outside the contact, since their effective shear stress  was found to be {5.8}~MPa at 20~m/s and 3.0~MPa at 2~m/s each time at $\Gamma=1$.
%
Thus, for islands with ten times the radius, areal friction can dominate 
but the friction stemming from non-contact is not yet necessarily negligible.
In fact, when the mean areal stress decreases with contact area, as is the case for structurally lubric contacts,  the non-contact friction may well be dominant, in particular at the small (or zero) normal loads studied in this work. 
%
Having provided expectations from atomistic simulations, we return to the presentation and interpretation of experimental results.

Previous studies show that a freshly prepared graphite surface becomes completely coated with molecular contaminants in just a few minutes when exposed to air \cite{li2016water}. 
Thus, considering that our experiments are conducted under uncontrolled, ambient conditions, some surface contamination will be present, even if it does not reveal itself in AFM images.
This is why we classify a sample as \textit{lightly contaminated} even if the contamination is not revealed by AFM. 
However, when contamination is clearly visible in phase and topography images,  which typically happens $>$1 week after synthesis, we deem the sample to be \textit{heavily contaminated}.
To explore the impact of the degree of contamination on aging, rejuvenation, and friction switches, we repeated our nanomanipulation experiments for \textit{heavily contaminated} islands.
Table~\ref{table:quantitative_new} reports the results.
The numbers expressed under ``frac." (short for fraction) contain the number of experiments in the denominator.
The fractions are converted to percentages for clearer comprehension.
Remarkably, we observe that increased levels of contamination significantly diminish the observation frequency of all three effects (rejuvenation, aging, and friction switches), by 75\% and more.


    

\begin{table} [ht]
\centering
\small
\begin{tabular}{P{3cm}|P{0.8cm}P{0.8cm}|P{0.8cm}P{0.8cm}|P{0.8cm}P{0.8cm}}
\hlineB{2.5}
{Degree of} 
& \multicolumn{2}{c|}{\it Rejuvenation} &  \multicolumn{2}{c|}{\it Aging} & \multicolumn{2}{c}{\it Switches} \\ 
%
Contamination 
& frac. & \% & frac. & \% & frac. & \%\\
\hline
{light} & 26/26 & 100 &  8/\phantom{2}8 & 100 & 15/26 & 58\\ 
{heavy} & 20/82 & \phantom{1}{24}  & 5/22 & \phantom{1}{23}  & \phantom{1}6/82 & \phantom{5}{7} \\ 
\hlineB{2.5}
    
\end{tabular}
\caption{Comparative analysis of observation frequencies of the three effects for lightly contaminated and heavily contaminated islands}
\label{table:quantitative_new}
\end{table}

\begin{figure}
\centering
\includegraphics[width=1\textwidth]{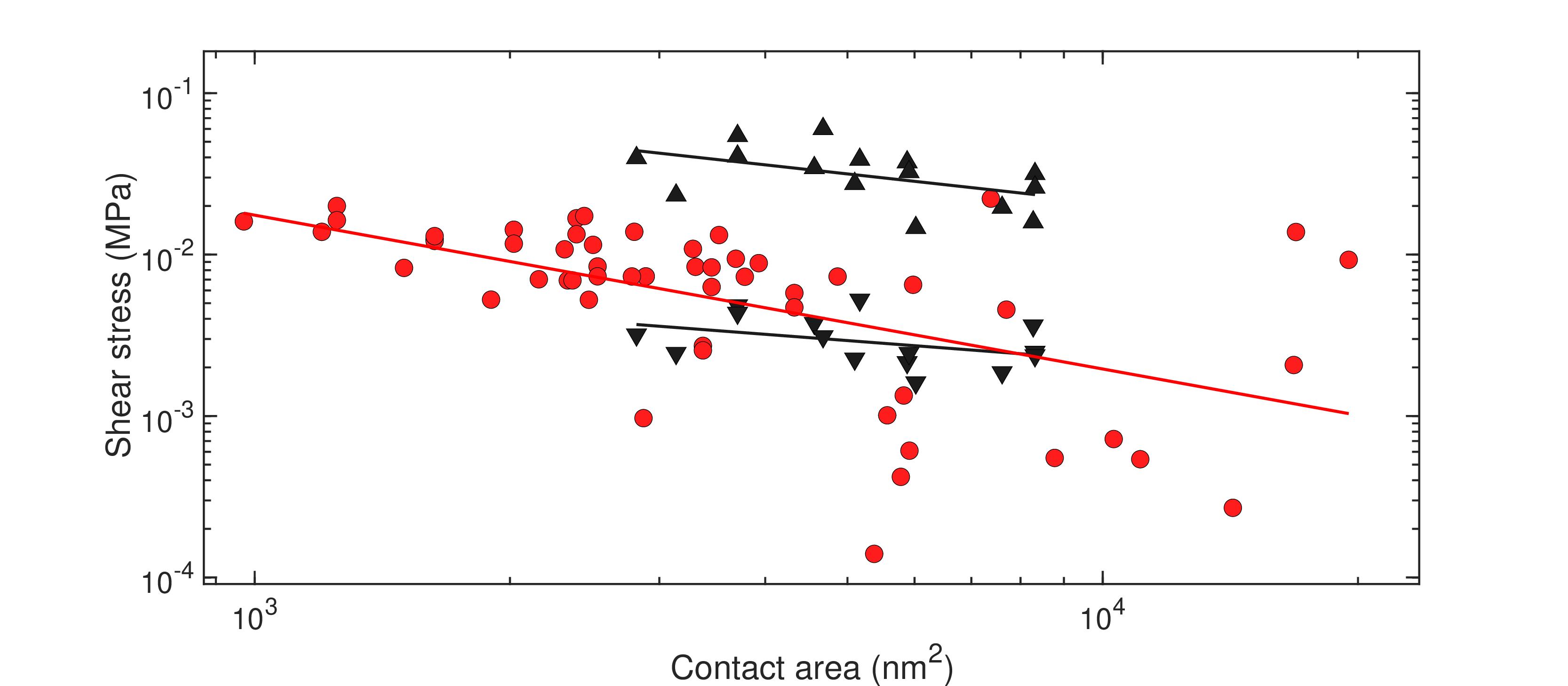}
\caption{Shear stress as a function of contact area recorded for lightly contaminated and heavily contaminated islands. Black triangles represent shear stresses associated with lightly contaminated islands that exhibit the switch effect, separated into two branches (upward triangles for the high friction branch, downward triangles for the low). Red circles represent shear stresses associated with heavily contaminated islands that do not exhibit any of the three effects.}
\label{fig_stressvsarea}
\end{figure}

While all islands in lightly contaminated samples exhibit the rejuvenation effect and more than half the switch effect (with friction between the two branches differing by factors from three to ten), the majority of heavily contaminated islands show consistent friction levels across all manipulations. 
The associated shear stresses align more closely with the low-friction branch of the lightly contaminated islands, with the exception of a few data points, as can be seen in Figure~\ref{fig_stressvsarea}, which shows shear stresses as a function of contact size $A$. Please note that Figure~\ref{fig_stressvsarea} specifically contains data for lightly contaminated islands that exhibit the switch effect (15), and for heavily contaminated islands that do not exhibit any of the three effects (55). 
All three cases, i.e., heavy contamination as well as high and low friction branches at light contamination, reveal the trend that larger islands have smaller shear stresses.
This observation can be a sign of structural lubricity and/or a friction mechanism that is dominated by the contact line. 
Surprisingly, the trend is most pronounced in the case of heavy contamination, where $\tau$ evolves almost linearly with $1/A$, at least for $A \lesssim 15,000$~nm$^2$. 
Such strong scaling of the shear stress does not appear plausible outside the realm of structural lubricity. 

The low friction branch shows the weakest scaling, i.e., $\tau \propto A^{-0.4}$. 
Given the scatter in the data, this result is consistent with friction that is dominated by contact line effects, i.e., the $\tau(A)$ dependence is close to a $\tau \propto 1/\sqrt{A}$ power law. 
The shear stress of $\mathcal{O}(20~\text{kPa})$ matches previous simulation results by two of us on gold islands sliding past perfectly clean graphite substrates~\cite{Gao2022FC}, after linear extrapolation to $v = 0$.
Thus, there is the possibility that the 
the low-friction branch corresponds to the sliding of gold islands on perfectly clean graphite.
However, for this to be true, the velocity dependence would have to be rather weak.
This could be the case, because changes in the Moiré-type patterns produced by normal displacements indicated the friction to be driven by instabilities, which generally induces rather weak, i.e., logarithmic-type velocity dependence of friction.~\cite{Muser2011PRB}
These corrections cross over to a steep linear dependence at a very small velocity~\cite{Muser2011PRB}.
Thus, we would expect an extrapolation of the sliding of gold cluster on a perfectly clean substrate to produce even smaller shear stresses than those measured experimentally. 
%
%
%
%
We  conclude that the experimental surfaces should still be covered by a thin, well-ordered layer of adsorbates.
%
This additional mono-layer (or potentially, double-layer) increases the friction compared to clean graphite-gold contacts, albeit not to the extent that superlubricity is destroyed. 

The high-friction branch shows scaling of roughly $\tau \propto {A}^{-0.6}$, which is again close to $\tau \propto 1/\sqrt{A}$.
One possible explanation would be that there are partial contamination layers at the early stages after sample preparation. 
Moving and thus removing these partial layers costs much energy, as has become evident from the MD simulations.
%
%
For a thick contamination layer, i.e., thicker than the correlation length of the density-density autocorrelation function, discrete steps from one-to-two or two-to-three monolayers no longer occur, and a sliding tip no longer removes fractional monolayers from the surface. 
Another possibility for the disappearance of the effect of fractional layers after very long waiting times could be that fractional layers are far from thermodynamic equilibrium, which is reached after long waiting times. 

Besides molecular contamination, various other factors could contribute to the noted friction duality within the structural lubricity regime, including the presence of dislocations \cite{sharp_elasticity_2016} and structural defects \cite{minkin_atomic-scale_2021} in the islands. 
However, shear stresses in the experiments are rather below than above those in the simulations so that we do not see the need to introduce additional effects into the simulations that would further increase that gap.
Moreover, during the initial stages of our simulations on this topic, i.e., before introducing contamination, we explored gold clusters with point and line defects.
Yet, friction remained much below that observed with contamination. 
Potentially more importantly, the von Mises yield strength of gold at the macroscopic scale is on the order of 100~MPa, while expertimental shear stresses are clearly less than 0.1~MPa at the nanoscale, at which point introducing dislocations becomes difficult or rather impossible. 
Lastly, we remind the reader of the mild, zero/negligible load condition used experimentally, which excludes the possibility of structural defects in graphite becoming relevant.


In summary, we find that gold islands on graphite under ambient conditions and driven under negligibly small normal load exhibit substantial increases in friction after waiting times ranging from 30~minutes to a day, \textit{aging}, dramatic friction reduction after the first areal scan, \textit{rejuvenation}, but also spontaneous jumps between low and high friction domains during a given scan, \textit{friction switches}.
The magnitude of the aging and rejuvenation was typically on the order of three to ten, which substantially exceeds the usual 10\%-30\% relative changes reported for these phenomena in both macroscopic and single-asperity contacts. 
Moreover, friction changes are reversible (via \textit{switches}) and thus differ from situations like the onset of catastrophic wear or permanent material damage. 
They also differ from previously reported \textit{frictional duality}~\cite{dietzel_frictional_2008}, which was related to the difference between perfectly clean and contaminated surfaces, rather than differing friction levels for different degrees of contamination. 
All three effects are diminished and shear stresses are small after waiting times exceeding one week, i.e., when surfaces are heavily contaminated with air-borne molecules. 

Since (1) both graphite and gold are chemically inert, (2) the gold islands are small and driven at negligibly-small normal loads, and (3) resulting shear stresses are well below 0.1~MPa, we can rule out dislocations, chemical defects, bond formation or other mechanisms requiring either large stresses or large energy densities.
As such, we see contamination as the main suspect that causes the observed phenomena. 
However, it is not straightforward to understand why initial contamination after minutes to hours exposure of graphite to ambient conditions causes high friction, while very long exposure causes small friction, in particular in light of the rejuvenation effect, which we believe to be caused by the removal of contaminants. 

Thus, to elucidate the role of contaminants in our experiments, we simulated gold islands on graphite with different levels of contamination.
The MD simulations reveal a highly non-monotonic dependence of the shear stress on coverage density.
Gliding over sparse mono- and double-layers show low shear stresses, while dense ones lead to intermediate stresses.
The largest shear stresses are observed when the island is in direct contact with the substrate but ``plows'' through and thus potentially removes a sub-monolayer of (highly-ordered) adsorbed molecules.
These observations can be in line with the experimental results, if we assume that friction of a slightly contaminated scan is high during the first areal scan, if this first scan removes partial layers, and low during any subsequent scan as long as no new partial layer has condensed onto the surface. 
However, this interpretation is admittedly speculative in nature, given the large gap in velocity and (linear) size between simulations and experiments, which are nine orders of magnitude and a factor of 10 to 30, respectively. 
Extrapolating results are difficult for a variety of reasons.
Most importantly,  friction-velocity scaling relations generally have a limited range of validity.
Typically, the response becomes more Stokes-like as velocity decreases, in which case shear-thinning effects are reduced at small velocity.
However, even simple polymeric systems can show the opposite trend~\cite{Mees2023JCP}.
Moreover, there will not be a single dissipation channel.
The relative contribution from shear stresses caused by the motion of the contact line or by dynamics caused outside the contact will be larger for the small \textit{in-silico} islands 
{than} the large, \textit{real} islands. 

Despite the difficulties of directly relating simulation to experimental results, the simulations clearly reveal a dependence of shear stresses on the relative orientation between the gold island and graphite at one- to two-monolayer coverage. 
This does not only explain the large experimental scatter but also reveals that superlubricity is still in effect despite the presence of contamination and even for islands with radii as small as 3~nm.
The conclusion from the simulations that gold islands can slide in superlubric fashion on graphite, even under lightly and heavily contaminated conditions caused by adsorbed contaminant molecules, finally provides an answer as to why ultra-low friction was observed experimentally for this material system under ambient conditions \cite{cihan_structural_2016}. This is critical as the experimental observation of structural lubricity under ambient conditions came as a surprise to the scientific community, where, until that point, it was assumed that molecular cleanliness was a prerequisite for the phenomenon. Our conclusions can also be potentially extended to explain the ultra-low friction exhibited by turbostratic graphite under ambient conditions~\cite{Kumar2011TI, Morstein2022NC}, due to the finding of ultra-low shear stresses even under the presence of adsorbed alkanes and for small contact areas a few nanometers in lateral span.
%

\begin{acknowledgement}

This work was supported by the National Science Foundation (NSF) via award number 2131976 and the German Research Foundation (DFG) under grant number GA 3059/2-1. W.H.O. acknowledges support from the NSF in the form of a Graduate Research Fellowship (GRF). The samples studied here were synthesized at the Molecular Foundry, Lawrence Berkeley National Laboratory. Work at the Molecular Foundry was supported by the Office of Science, Office of Basic Energy Sciences, of the U.S. Department of Energy under Contract No. DE-AC02-05CH11231.

\end{acknowledgement}

\FloatBarrier


\end{document}